\documentclass[a4paper,reqno,nofootinbib,superscriptaddress,12pt]{revtex4}
\usepackage{tikz}
\usetikzlibrary{decorations.pathmorphing,patterns}
\usepackage{graphicx,psfrag}
\usepackage{dcolumn}
\usepackage{epsfig}

\usepackage[centertags]{amsmath}
\usepackage{amsfonts,amsmath,color}
\usepackage{euscript}
\usepackage{amssymb}
\usepackage{amsthm}
\usepackage{newlfont}
\usepackage{mathrsfs}
\usepackage{subfigure}

\newcommand{\opunit}{\text{1}\kern-0.22em\text{l}}

\newcommand{\id}{\textrm{d}}

\newcommand{\R}{\mathbb R}

\newcommand{\beq}{\begin{equation}}
\newcommand{\eeq}{\end{equation}}

\newcommand{\Oe}{{+\mathcal{O}(\varepsilon^3)}}

\usepackage{ulem}

\def\bea{\begin{eqnarray}}
\def\eea{\end{eqnarray}}
\def\ba{\begin{array}}
\def\ea{\end{array}}

\def\la{\langle}
\def\ra{\rangle}
\def\del{\partial}

\begin{document}

\title{Frenetic aspects of second order response}
\author{Urna Basu}
\email{urna.basu@fys.kuleuven.be}
\affiliation{Instituut voor Theoretische Fysica, KU Leuven, Belgium}
\author{Matthias Kr\"uger}
\email{mkrueger@is.mpg.de}
\affiliation{4th Institute for Theoretical Physics, Universit\"at Stuttgart, and Max Planck Institute for Intelligent Systems, Stuttgart, Germany}
\author{Alexandre Lazarescu}
\affiliation{Instituut voor Theoretische Fysica, KU Leuven, Belgium}
\author{Christian Maes}
\affiliation{Instituut voor Theoretische Fysica, KU Leuven, Belgium}
\date{\today}

\begin{abstract}
Starting from second order around thermal equilibrium, the response of a statistical mechanical system to an external stimulus is not only governed by dissipation
and depends explicitly on dynamical details of the system. The so
called frenetic contribution in second order around equilibrium is illustrated in different
physical examples, such as for non-thermodynamic aspects in the coupling between system and
reservoir, for the dependence on disorder in dielectric response and for the nonlinear correction to the Sutherland--Einstein relation.  More generally, the way in which a system's dynamical activity changes by the pertubation is visible (only) from nonlinear response.
\end{abstract}
\maketitle

\section{Introduction}
The first order or linear response around thermodynamic equilibrium is statistically given in the fluctuation--dissipation theorem \cite{cal,kubo}.  
Its validity and meaning are basically unchanged in going from quantum to classical dynamics or in starting from Liouville or from Langevin descriptions: these Kubo and Green-Kubo formul{\ae} for linear response have a clear thermodynamic interpretation correlating the observable with the excess entropy flux due to the perturbation.  The present paper investigates what remains true of that for nonlinear response.\\
  As cannot be too surprising but much in contrast to linear response for equilibrium systems, nonlinear response depends directly on certain details of the dynamics.  One can thus not truly speak about \emph{the} second order response as would be determined thermodynamically. For example, thermodynamically equivalent perturbations which differ kinetically will yield different nonlinear responses. Similarly, for mechanical perturbations  again from second order response onwards the dissipated power appears correlated with changes in e.g. escape rates or in the local kinetic temperature.  The present paper offers a unifying framework for describing these effects.\\

Rather limited attention is  paid in the textbook literature to the systematics of extending linear response theory to stronger perturbations. Even second order response around equilibrium is hardly ever made explicit.  The explanation is probably that the usual more analytic approach starting from a Dyson series expansion fails to be very informative in general; see Appendix \ref{compu} for that standard response formalism.  Nevertheless there is a large research literature on second and nonlinear order response albeit mostly restricted to specific models or computational schemes.   References include \cite{lip,mik,mik2,parr,opp,dell,diez,val}.  The present paper takes a different perspective following Refs.~\cite{fdr,col}, unraveling the second order response in essential physical contributions that may indeed depend on dynamical details irrelevant for linear response.  Inversely, measuring nonlinear response will give kinetic information.\\
The next Section explains the main issue of the paper.   A path-integral approach appears particularly useful here, as will start in Section \ref{path} employing the concept of dynamical ensembles.  That indeed brings to the foreground a kinetic (in addition to the thermodynamic) dependence in the response formula, where the frenetic contribution for linear response around nonequilibrium conditions \cite{fdr,fren} gets an analogue in the second order response around equilibrium. We illustrate these frenetic aspects in a number of cases in Section \ref{ex}.  For studying a perturbation that modifies the coupling of a chemical reservoir with a particle system we use the open zero range process:  changing the chemical potential of the environment can be obtained in multiple kinetic ways which become visible (only) at second order in the perturbation.  For frequency-dependent dielectric response (change of polarization of a dielectric material in a periodic electric field) we use a lattice Lorentz model where an 
external 
field also changes the collision frequencies with the obstacles,  a time-symmetric effect which contributes in second order response.  A further example concerns the motion of colloidal particles in narrow channels where the current picks up a nonlinear response to a driving field
that also depends on the equilibrium correlation between current and local diffusion coefficient, or between current and residence times in local potential minima.  That gives the second order correction to the traditional Sutherland--Einstein relation between mobility and diffusion constant. 
Section \ref{irrt} considers the question in the context of irreversible thermodynamics which is however unable to give the response in terms of fluctuations.

\section{Introductory examples: What is at stake?}

From the possible perturbations of an open system in equilibrium with its environment, consider the exemplary case where particles can exit and enter the system and where the chemical potential of the environment is changed.  Clearly, the (directed) particle current into the environment will be affected for some moment, with an adjustment of the average number of particles in the system. That entropic aspect enters the linear response and embodies the fluctuation--dissipation theorem.  But, any change of chemical potential is felt in the system through a specific coupling with the environment.  There are in fact many ways in which the entrance  and exit rates $R_{\text{in}}, R_{\text{out }}$ can change separately while their ratio is adjusted to the new chemical potential $\mu_{\text{new}}$ via the condition of detailed balance,
\[
\frac{R_{\text{in}}}{R_{\text{out}}} =  e^{\beta\mu_{\text{new}}}
\]
where $\beta$ is the inverse temperature of the environment.  Depending on these kinetic factors, the perturbation in chemical potential produces changes in the (undirected) traffic of particles between system and reservoir. We call it excess dynamical activity and it will enter in the second order response.  The `frenetic aspects' in the title of this paper refers to that change in volatility or busy-ness of the system, in contrast with and complementing the more usual time-antisymmetric aspects.\\
Consider next a mechanical perturbation to the open system.  The extent to which an extra force is felt on the particles of the system obviously depends also on how their reactivity and diffusivity get changed under the force.  The mechanisms of how external stimuli may influence internal processes beyond linear response and indeed beyond purely dissipative effects is a question of much recent interest with many remaining puzzles. For example, extra mechanical stress on a cell can affect gene expression \cite{gen,gene}; hyper-gravity may change reactivities \cite{hyp}; under certain circumstances particle interaction can lead to jamming under applied fields \cite{jam}; return to a new equilibrium can be postponed for a very long time in the case of ageing \cite{age}, and response is not simply characterized via the fluctuation--dissipation theorem.\\
Nonlinear response is expected to pick up these finer issues.  What is at stake then is to see whether there still exists some systematics in dealing with higher order response (second order for the present paper).  Can one in other words obtain some more universal description in which the variety of kinetic changes is summarized into one or a few quantities, somewhat the analogue of entropy fluxes summarizing the dissipative effects?  Our answer here is yes {\it in theory},  which is the good news in view of the great many kinetic details that can be affected by a 
perturbation.   The point about theory is that one has to look at how the perturbation changes the time--symmetric component of the action of the dynamical ensemble describing the reduced dynamics of the open system. That is made clear in the next Section \ref{path}.  We thus take here the point of view of theoretical physics and less that of computational physics to create a framework and to identify explicit fluctuation and correlation functions that describe the measurable response.  These involve frenetic aspects as announced in the title, which means that they pertain to changes in expected dynamical activity.

\section{From path-integration to response formula}\label{path}
Dynamical ensembles express the weight or plausibility of a trajectory of system variables. Let $\omega$ denote such a system trajectory or path on some reduced level of description and over time-interval $[0,t]$.
Take a path observable $O=O(\omega)$; its expectation is given by the path-integral
\begin{equation}\label{sta}
\langle O \rangle = \int {\mathbb D}[\omega] \, P(\omega)\,O(\omega) =  \int {\mathbb D}[\omega] \,e^{-\cal A(\omega)}\,P_{\text{eq}}(\omega)\,O(\omega)
\end{equation}
where ${\mathbb D}[\omega]$ is the formal (and further unimportant) volume element on path--space.
Here the perturbed dynamical ensemble $P$ is characterized by an action $\cal A$ with respect to the reference equilibrium ensemble $P_{\text{eq}}$. In other words, we assume there is a density $\exp[-{\cal A}]$ on path-space of the probability distribution $P$ with respect to the reference equilibrium $P_{\text{eq}}$,
\[
P(\omega) = e^{-\cal A(\omega)}\,P_{\text{eq}}(\omega).
\]
At time zero the system is described by the equilibrium distribution on the system states and the path-probability distributions $P$ and $P_{\text{eq}}$ differ because $P$ takes into account the perturbation.  We consider here the case where the system keeps weak contact with an equilibrium environment.  Reversibility of the equilibrium condition demands that $P_{\text{eq}}(\theta\omega) = P_{\text{eq}}(\omega)$ for time-reversal $\theta$ defined on paths $\omega = (x_s, 0\leq s\leq t)$ via  
\[
(\theta\omega)_s =\pi x_{t-s}
\]
 for kinematical time-reversal $\pi$ (e.g. flipping the sign of velocities), which in practice expresses itself as the condition of (possibly generalized) detailed balance for the unperturbed dynamics. \\

  The perturbation such as adding an extra potential or nonconservative force of strength $\varepsilon \ll 1$ to model the external stimulus is present in the action ${\cal A}$. We assume that ${\cal A} = {\cal A}_\varepsilon$ depends smoothly on the small parameter $\varepsilon$, i.e., ${\cal A}_\varepsilon={\cal A}_0+{\cal A}'_0\varepsilon+\frac{1}{2}{\cal A}''_0\varepsilon^2 \dots$, with ${\cal A}_0=0$. We also take here the case where the perturbation is time-independent (after the initial time); the case of applying a perturbation for a general time--protocol is postponed to Appendix \ref{genca}.\\

 It turns out to be useful here to decompose the resulting action $\cal A$ into a time-antisymmetric $S(\omega)$ and a time-symmetric $D(\omega)$ term:  $\cal A = D-S/2$ with $S\theta=-S$ and $D\theta=D$ .  These are called, for good reasons \cite{timer,fdr}, the  entropic and  frenetic components of the action and they are of course functions of the type of perturbation. We will restrict to perturbations for which $\varepsilon$ enters at most to linear order in $S$, i.e., its second derivative $S''_{\varepsilon=0} =0$. This includes most of the relevant physical examples, e.g., for perturbations via potential forces, it means that the Hamiltonian is linear in the perturbing field. This specifies the order of the perturbation. Note that $S(\omega)$ and $D(\omega)$ depend on time $t$ because they are defined on paths $\omega$ in the time-interval $[0,t]$.\\

We expand \eqref{sta} to second order around equilibrium as 
\begin{equation}\label{2n}
\langle O \rangle = \langle O \rangle_{\text{eq}} - \varepsilon \,\langle {\cal A}'_0(\omega)\, O(\omega)
\rangle_{\text{eq}} - \frac{\varepsilon^2}{2}[\langle {\cal A}''_0(\omega)\, O(\omega)\rangle_{\text{eq}} - \langle({\cal A}'_0(\omega))^2\, O(\omega)\rangle_{\text{eq}}] + \mathcal{O}(\varepsilon^3)
\end{equation}
with the right--hand side in terms of expectations
\[
\langle F(\omega)\rangle_{\text{eq}} :=  \int {\mathbb D}[\omega] \,P_{\text{eq}}(\omega)\,F(\omega)
\]
in the equilibrium process.
The same expression as \eqref{2n} for the time-reversed observable reads
\begin{equation}\label{2n2}
\langle O \theta\rangle = \langle O \rangle_{\text{eq}} - \varepsilon \,\langle {\cal A}'_0(\theta\omega)\, O(\omega)
\rangle_{\text{eq}} - \frac{\varepsilon^2}{2}[\langle {\cal A}''_0(\theta\omega)\, O(\omega)\rangle_{\text{eq}} - \langle({\cal A}'_0(\theta\omega))^2\, O(\omega)\rangle_{\text{eq}}]\Oe
\end{equation} 
where we have used the time-reversal invariance of the equilibrium process, $\langle F(\omega)\rangle_{\text{eq}} = \langle F(\theta\omega)\rangle_{\text{eq}} $.
Subtracting \eqref{2n2} from \eqref{2n} gives
\[ 
\langle O - O\theta\rangle = -\varepsilon \,\langle [{\cal A}'_0(\omega)-{\cal A}'_0(\theta\omega)]\, O(\omega)
\rangle_{\text{eq}} + \frac{\varepsilon^2}{2} \langle[({\cal A}'_0(\omega))^2 - ({\cal A}'_0(\theta\omega))^2]\, O(\omega)\rangle_{\text{eq}}\Oe
\] 
where we already took ${\cal A}''_0(\theta\omega) = {\cal A}''_0(\omega)$ from our set-up where $S_\varepsilon = \varepsilon \,S'_0$. 
Plugging in the decomposition ${\cal A}_\varepsilon = D_\varepsilon-S_\varepsilon/2$ and using $D'\theta = D', S'\theta=-S'$, we find
\begin{equation}\label{2n4}
\langle O - O\theta\rangle = \varepsilon \,\langle S'_0(\omega)\, O(\omega)
\rangle_{\text{eq}} - \varepsilon^2\,\langle D'_0(\omega)\,S'_0(\omega)\, O(\omega)\rangle_{\text{eq}}\Oe
\end{equation}
 where $D'_0,S'_0$ are first derivatives evaluated at $\varepsilon =0$.\\
For a state observable $O(\omega)=O(x_t)$ we have $O\theta(\omega)= O(\pi x_0)$ and we can use $\langle O(\pi x_0)\rangle_{\text{eq}} = \langle O(x_0)\rangle_{\text{eq}} = \langle O(x_t)\rangle_{\text{eq}}$.  Applying formula \eqref{2n4} to that case we obtain the extension  to second order of the traditional Kubo formula,
\begin{equation}\label{kubo2}
\langle O(x_t)\rangle - \langle O(x_t)\rangle_{\text{eq}} =  \varepsilon\,\langle S'_0(\omega)\,O(x_t)
\rangle_{\text{eq}} - \varepsilon^2\,\langle D'_0(\omega) \,S'_0(\omega)\, O(x_t)\rangle_{\text{eq}}\Oe.
\end{equation}
We will prove in Appendix \ref{genca} that Eq. \eqref{kubo2} holds for general time-dependent
perturbation protocols.\\
When on the other hand we have $O(\theta\omega) = -O(\omega)$ expressing oddness under time-reversal, as for time-integrated particle or energy currents $O(\omega) = J(\omega)$, we get from \eqref{2n4}  the extended Green--Kubo formula
\begin{equation}\label{gk}
\langle J\rangle = \frac{\varepsilon}{2} \,\langle S'_0(\omega)\, J(\omega)
\rangle_{\text{eq}} - \frac{\varepsilon^2}{2}\,\langle D'_0(\omega)\,S'_0(\omega)\, J(\omega)\rangle_{\text{eq}}\Oe.
\end{equation}
The linear order in \eqref{kubo2} or in \eqref{gk} is expressed as a correlation between observable and the entropic component $S_\varepsilon$ only.  That is the essence of the fluctuation--dissipation relation, and below more explicit versions will appear so that the relation with (Green-)Kubo relations will become plain. 
When $S'_0=0$ there is no linear and no second order response.  On the other hand, when the frenetic component $D$ is constant (and hence unchanged under a small perturbation), there is \underline{no second order} response at all.  (The reverse is not true). The purpose of the present paper is to investigate the meaning and consequences of that dependence  on $D'_0$ in second order.\\
To complete the discussion, we remark that for time-symmetric observables the linear response already picks up the dynamical activity.  We see that by adding \eqref{2n2} and \eqref{2n}:
\begin{equation}\label{2n8}
\frac 1{2}\,\langle O + O \theta\rangle = \langle O \rangle_{\text{eq}} - \varepsilon \,\langle D'_0(\omega)\, O(\omega)\rangle_{\text{eq}} + O(\varepsilon^2).
\end{equation} 
That is relevant when considering for example a momentum current, which is time-symmetric. The second order expression for this case \cite{tin} is not considered here explicitly, although of course contained in Eq.~\eqref{2n}.

   Before we continue with further details on the response formul{\ae}, let us visit the above actors $D'_0$ and $S_0' $ more explicitly for three main classes of Markov dynamics \cite{kamp}.  In all cases we start from a detailed balance dynamics (or, for inertial dynamics --- generalized detailed balance) that determines the equilibrium distribution $P_{\text{eq}}$.\\

\subsection{Jump processes}\label{jup}
Consider a finite number of states $x,y,\ldots$ denoting for example energy levels, particle numbers or molecular configurations. We start from a system described by transition rates $k_{\text{eq}}(x,y)$ that satisfy detailed balance
\[
k_{\text{eq}}(x,y)\,\rho_{\text{eq}}(x) = k_{\text{eq}}(y,x)\,\rho_{\text{eq}}(y)
\]
with equilibrium distribution $\rho_{\text{eq}}(x)= e^{-\beta U(x)}/Z_\beta$ for energy $U(x)$ being exchanged with the heat bath at inverse temperature $\beta$.  The latter also describes the initial condition after which we are using perturbed transition rates, at times $s>0$,
\begin{equation}
k(x,y)= k_{\text{eq}}(x,y)\,[1+\varepsilon\phi(x,y)+ \frac{\varepsilon}{2}a(x,y) + O(\varepsilon^2)]\label{pa}.
\end{equation}
The parameterization distinguishes the time-symmetric component $\phi(x,y)=\phi(y,x)$  from the antisymmetric $a(x,y) = -a(y,x)$ in the  expansion to first order in parameter $\varepsilon\ll 1$.
The case of time-dependent $a_s(x,y)$ and $\phi_s(x,y)$ is treated in Appendix \ref{genca}.\\

 The entropic component in the response \eqref{2n4} or \eqref{kubo2} always equals
\begin{equation}\label{so}
S'_0(\omega) = \sum_{s_i} a(x_{s_{i^-}},x_{s_i})
\end{equation}
with the sum over jump times $s_i$ when $x_{s_{i^-}}\rightarrow x_{s_i}$ as dictated by the path $\omega$  (we use $i^-\equiv i-1$).\\
A commonly considered  perturbation makes a potential change $U\rightarrow U -\varepsilon\,V$ for which
\begin{equation}\label{ss}
a(x,y) = \beta
 \,[V(y) - V(x)] \;\implies\; S'_0(\omega) =\beta[V(x_t)-V(x_0)]
\end{equation}
which is heat over temperature,  or indeed the excess entropy flux (per $k_B$) to the heat bath during the trajectory $\omega$ due to the potential perturbation.\\

The way in which reactivities change under the perturbation, or in other words, how $\phi(x,y)$ looks like for a potential perturbation is in principle given from reaction rate theory by a version of the Arrhenius--Kramers formula \cite{shi}. It will not be reproduced here.\\ 
Continuing with  the parameterization \eqref{pa} we find the frenetic component
\bea
D'_0(\omega) = -\sum_{s_i} \phi(x_{s_{i^-}},x_{s_i}) + \int_0^t\id s \sum_y k_{\text{eq}}(x_s,y)\,[\phi(x_s,y) + \frac 1{2} a(x_s,y)] \label{eq:Dprime}
\eea
where the first sum is again over the jump times in the path over $[0,t]$.  That is then the quantity that crucially enters the second order (but not the linear) response \eqref{kubo2}  or \eqref{gk}.  We call 
\eqref{eq:Dprime} an excess dynamical activity as it involves the time-integrated change in escape rates, together with the sum of excess reactivities for all transitions in the path. Such a concept plays an increasing role in studies of nonequilibrium statistical mechanics; see e.g. \cite{vW,chan,fren} for some examples.

\subsection{Overdamped diffusion}\label{od}

For position $x\in \R^d$ an overdamped diffusion is defined in the It\^o-sense by
\begin{equation}
\label{overd}
\dot x_s = \chi(x_s)\,f(x_s) + \nabla {\cal D}(x_s) +\sqrt{2{\cal D}(x_s)}\,\xi_s .
\end{equation}
The environment is taken at inverse temperature $\beta$ so that $\chi(x) =\beta {\cal D}(x)$
for the mobility $\chi$ with respect to the diffusion matrix ${\cal D}$; they are strictly positive (symmetric)
$d \times d$-matrices. The process $\xi_s = (\xi_s^i, i=1,\ldots,d)$ is  standard white noise, i.e., $\langle \xi_s^i\rangle =0$ and $\langle \xi_s^i \xi_{s'}^j\rangle = \delta_{ij}\,\delta(s-s')$.  The $f$ collects all the forcing, including via confining potentials.  The equilibrium process has only conservative forces $\propto \nabla U(x)$ independent of time for which the equilibrium dynamics is reversible.  We  assume of course that $U(x)$ grows sufficiently at infinity.
For perturbation we consider a potential $V$ entering $f(x) = -\nabla U(x) + \varepsilon \,\nabla V(x)$ with a small amplitude $\varepsilon$ non-zero only for positive times $s>0$.  In contrast with the previous jump--case we do \underline{not} consider here the case that $\chi$ (or ${\cal D}$) would also change by the addition of that potential $V$.  Nevertheless the observables in the correlation function for the second order response will now involve separately the mobility $\chi$ and the diffusion ${\cal D}$, instead of having them only in the thermodynamic ratio $\chi/{\cal D} =\beta$ as in linear response.\\

It is easy to find again  that as in \eqref{ss}
\begin{equation}\label{sd}
S'_0(\omega) =\beta[V(x_t)-V(x_0)]
\end{equation}
while the frenetic component  equals
\begin{equation}\label{16}
D'_0(\omega) = {\beta\over 2}\,\int_0^t\id s\,  [-\chi\nabla U\cdot \nabla V + \nabla\cdot ({\cal D}\nabla V)](x_s) . 
\end{equation}
The differential operator between $[\cdot]$ is the backward generator $L$ of the equilibrium process:  
\[
D'_0(\omega) = {\beta\over 2 }\,\int_0^t\id s\, LV(x_s),\quad  LV(x) = \frac{\id}{\id s} \langle V(x_s)|x_0=x\rangle_{\text{eq}}(s=0).
\]
Here, $\langle V(x_s)|x_0=x\rangle_{\text{eq}}$ denotes the equilibrium average of $V(x_s)$ given the initial configuration $x_0=x$.

\subsection{Inertial Langevin dynamics}
For a single underdamped particle in a heat bath at temperature $T=\beta^{-1}$ we write the Langevin equation for position and velocity,
\bea
\dot x_s &=& v_s  \cr
m \dot v_s &=& - \gamma v_s   -\nabla U(x_s)  + \sqrt{2 \gamma T}~ \xi_s + \varepsilon\, \nabla V(x_s)\label{eq:in}
\eea
where the $\varepsilon-$perturbation is switched on at $s=0$.  Compared with \eqref{overd} we have $\gamma= 1/\chi$.  We now ignore any spatial dependence in the friction $\gamma$ for simplicity, and $\xi_s$ is distributed as in Eq.~\eqref{overd}.  Again, simple path-integral techniques \cite{fdr2} yield
the excess entropy flux
\bea
S_0'(\omega) = \frac 1{T}[V(x_t)-V(x_0)]\label{eq:16}
\eea
while the excess dynamical activity is time-symmetric and equals
\bea\label{177}
D_0'(\omega) = - \frac 1{2 \gamma T} \int_0^t  \nabla V(x_s)\, \big(m \frac{\id v_s}{\id s} + \nabla U(x_s)\big) \id s. 
\eea
The Langevin force $m \frac{\id v_s}{\id s} + \nabla U(x_s) = - \gamma v_s   + \sqrt{2 \gamma T}~ \xi_s$ appears here. For the connection with the overdamped case, note that the second term is the first term in \eqref{16} with the identification $\chi = 1/\gamma.$  
Partial integration of the first term in \eqref{177} leads to
\[
\frac{1}{2 \gamma T} \int_0^t \id s ~  \nabla^2 V mv_s^2 
- \frac{m h_t}{2\gamma T} \nabla V(x_t) v_t.
\]
The Smoluchowski (overdamped) limit is approached by identifying  $m v_s \to 0$  and $mv_s^2 = T$ on the Brownian time scale, which keeps only the first term in the above line and recovers the second term in \eqref{16} for $\gamma \,{\cal D} = T$.\\ 

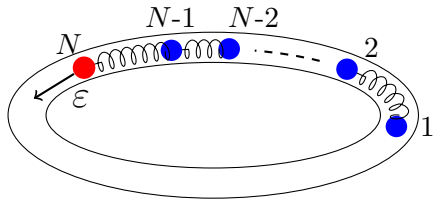
\begin{figure}
\centering
\begin{tikzpicture}
\draw (0,0) ellipse (2.2 cm and 0.7 cm);
\node[circle,fill=blue,inner sep=1mm] at (2.41,-0.15) {};
\draw[decoration={aspect=-0.5, segment length=1.5 mm, amplitude=1.2 mm,coil},decorate]  (2.45,0)  arc (0:40: 2.45 cm  and 0.88 cm); 
\node[circle,fill=blue,inner sep=1mm]  at (1.76,0.61) {};
\draw[dashed, thick]  (1.40,0.72)  arc (55:77: 2.45 cm  and 0.88 cm);
\node[circle,fill=blue,inner sep=1mm] at (0.21,0.88) {};
\draw[decoration={aspect=-0.5, segment length=1.5 mm, amplitude=1.2 mm,coil},decorate]  (0.08,0.88)  arc (88:100: 2.45 cm  and 0.88 cm);
\node[circle,fill=blue,inner sep=1mm] at (-0.55,0.86) {};
\draw[decoration={aspect=-0.5, segment length=1.5 mm, amplitude=1.2 mm,coil},decorate]  (-0.63,0.85) arc (105:130: 2.45 cm  and 0.88 cm);
\node[circle,fill=red,inner sep=1.05mm] (d) at (-1.7,0.63) {};
\draw[thick,->](d) -- (-2.36,0.22);
\draw (0,0) ellipse (2.7 cm and 1.1 cm);
\node at (2.82,-0.15) {$1$};
\node at (2.08,0.89) {$2$};
\node at (0.53,1.3) {\small $N$-$2$};
\node at (-0.56,1.3) {\small $N$-$1$};
\node at (-1.9,0.94) {\small $N$};
\node at (-1.74,0.22) {\large$\varepsilon$};
\end{tikzpicture}
\caption{Schematic representation of a polymer in a toroidal trap. The last monomer is being pulled by an additional force $\varepsilon.$\label{fig:poly}}
\end{figure}

Eq.~\eqref{eq:in} can as well be formulated for interacting particles, for which we take the example of a polymer of size $N$ in a toroidal trap (see Fig.~\ref{fig:poly}).  Each monomer $i=1,2,\ldots,N$ has position $q_s(i)\in S^1$ on the unit circle with corresponding (angular) momentum $p_s(i) = \dot{q}_s(i)$.  There is an  energy
\[
U(q,p) = \sum_{i=1}^N  \left\{\Phi(q(i)-q(i+1)) \right\}, \quad q(N+1)=0
\]
with coupling $\Phi$ between the monomers, an even and smooth function on the circle $S^1$.   The dynamics for $i=2\ldots, N-1$ is
\[
\dot p_s(i) = \Phi'(q_s(i-1)-q_s(i))-\Phi'(q_s(i)-q_s(i+1)) - \gamma p_s(i) + \sqrt{2\gamma T}\,\xi_s^i,
\]
for $i=0$,
\[
\dot p_s(0) = -\Phi'(q_s(0)-q_s(1)) - \gamma p_s(0) + \sqrt{2\gamma T}\,\xi_s^0
\]
and for $i=N$,
\[
\dot p_s(N) = \Phi'(q_s(N-1)-q_s(N)) + \varepsilon - \gamma p_s(N) + \sqrt{2\gamma T}\,\xi_s^N
\]
where the perturbation of strength $\varepsilon$ represents a constant driving force on the last monomer (and is not of potential type).  The temperature $T=\beta^{-1}$ is uniform and the $\xi_s^{i}$ are independent standard white noises (see below Eq.~\eqref{overd}). On the real line
there is of course no stationarity because there is no boundary, and the polymer would be driven wherever.  The toroidal set-up makes the perturbation non-conservative, yet stationary and may be advantageous for experimental studies: For $N=1$ the system was experimentally studied for its nonequilibrium response in Ref.~\cite{ser}. The corresponding dynamical time-reversal is $(\theta \omega)_s = (q_{t-s},-p_{t-s})$ for $0\leq s\leq t$; note the flip of momenta.  As a result, the dynamics is (generalized) reversible at $\varepsilon=0$ in the equilibrium process for stationary density $\rho_{\text{eq}}\propto \exp[-\beta (U + \sum_{i=1}^N \frac 12 p(i)^2)]$ with respect to $\prod\id p(i)\id q(i)$. \\

To find the action ${\cal A}$ defined in \eqref{sta}, we can use the procedure of \cite{heatcond}.  The result is that
\begin{equation}\label{sde}
S'_0(\omega) =\frac 1{T} \int_0^t p_s(N)\,\id s
\end{equation}
which  quantifies the time-integrated dissipated angular power per $\varepsilon$
(still independent of $\gamma$). The frenetic component equals
\[
D'_0(\omega) = {1\over 2\gamma T}\,\{p_t(N)-p_0(N) - \int_0^t\id s\, \Phi'(q_s(N-1)-q_s(N)) +\text{constant})\} 
\]
where the $\gamma-$dependence and the details of the interaction (coupling of polymer with its last edge) are apparent.
These could not be identified in linear response.

\section{Explicit Examples}\label{ex}

In this section we illustrate the frenetic aspects of second order response in greater details with the help of several explicit examples.

\subsection{Zero range process}

Our first  example is a model of a granular gas in one dimension, called the zero range process.  It is an exactly solvable  model where particles on a lattice hop to neighbouring sites at a rate which depends only on the departure site \cite{zrp}. The system is coupled to an environment thermodynamically characterized  via a given chemical potential.
Yet the more detailed coupling with the environment matters for response to second order:   there are different kinetic ways to change the chemical potential and, as we will see below, they make different second order responses. 
\begin{figure}[t] 
 \centering
 \includegraphics[width=10 cm,bb=0 0 496 210]{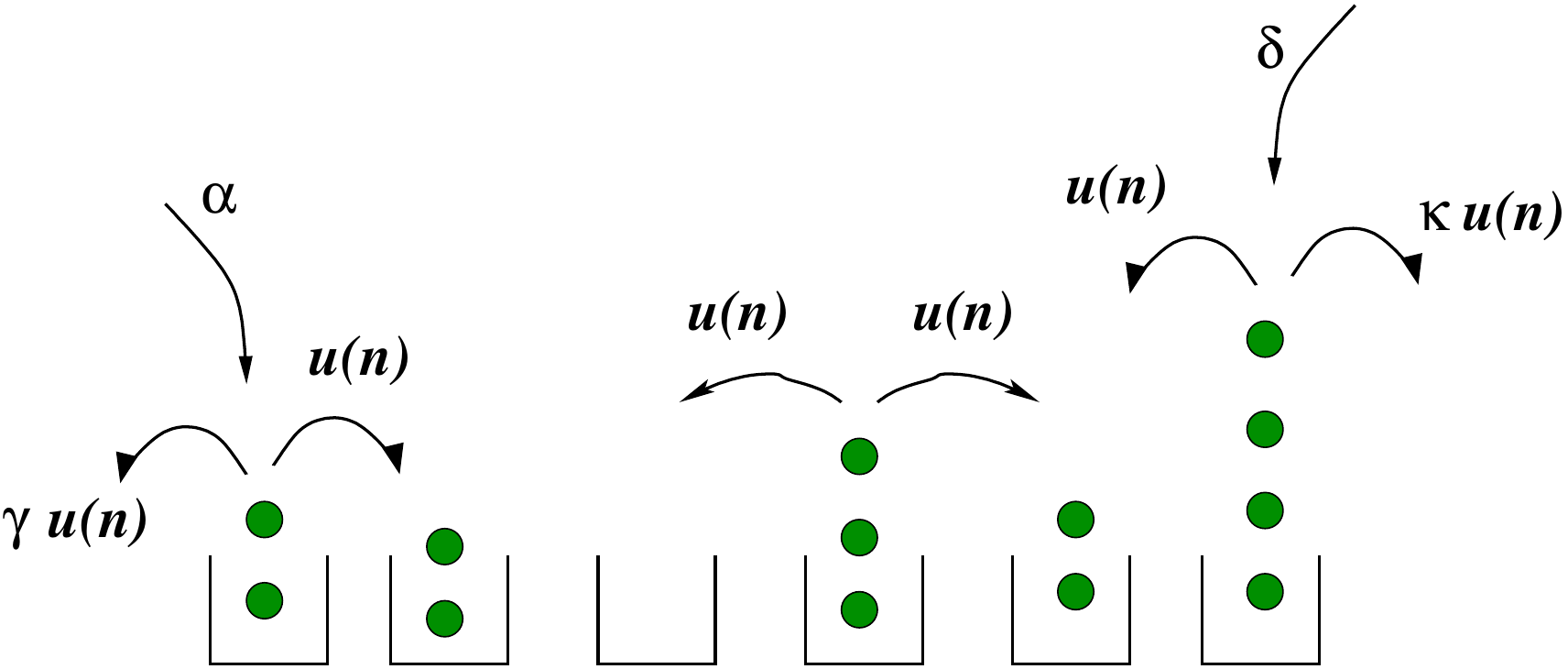}
 \caption{ Schematic representation of the open zero range process with respective transition rates as they depend on the local particle number $n$.  The chemical potential $\mu$ of the environment is given in \eqref{chemp} but does not fully determine the entry and exit rates.}
 \label{fig:zrp}
\end{figure}

The one--dimensional zero range process is defined on a lattice with $L$ sites; each site $i$  is occupied with $n_i$ number of particles. A particle from a bulk site $i$ hops randomly to its right or left neighboring site with rate $u(n_i)$.
In addition particles may enter at the left and right boundary sites $i=1$ and $i=L$ with rates $\alpha$ and $\delta$ respectively. A particle can also leave the system from the boundaries  with rate $\gamma\, u(n_1)$ from the first site and with rate $\kappa\, u(n_L)$ at the last site.  The equilibrium situation corresponds to $\alpha\kappa= \gamma \delta$ following the interpretation of detailed balance where the boundary rates correspond to the chemical potential $\mu$ and inverse temperature $\beta$ of the environment via
\bea\label{chemp}
\alpha/\gamma =  \delta/\kappa = e^{\beta \mu}.
\eea
Clearly however, the jump rates $\alpha,\delta,\gamma$ and $\kappa$ are not completely specified by $\mu\beta$; the coupling with the environment indeed depends on more kinetic elements.\\

Let us consider two ways of changing $\mu\rightarrow \mu+ \varepsilon $.

{\bf Case I.} First we consider the case when only the entry rates at both edges are perturbed $\alpha \to \tilde \alpha = \alpha e^{\beta \varepsilon}, \delta \to \tilde \delta = \delta e^{\beta \varepsilon}$. That gives an increase in the chemical potential to $\tilde{\mu}=\mu + \varepsilon.$ 
The entropic and frenetic components of the response are obtained from \eqref{so} and \eqref{eq:Dprime}:
\bea\label{34}
S_0^\prime &=& \beta J_{\text{in}} ,\cr
D_0^\prime &=& \beta (\alpha+\delta) t -\frac \beta 2 \,I . 
\eea
Here $J_{\text{in}}$  is  the net number of particles transferred into the system from the environment during the time interval $[0,t]$ and $I$ is the total number of particle exchanges between the system and the reservoir during the same time. We see that while the entropic component is completely specified by $\beta$ and the current, the frenetic component explicitly depends on the entrance rate $\alpha+\delta$ and introduces the time-symmetric ``traffic'' $I$.\\
The second order response inserts \eqref{34} into \eqref{kubo2},
\bea
\la S_0'(\omega ) D_0'(\omega ) O(t) \ra_{\text{eq}} &=& \beta^2 \left\la J_{\text{in}} \left[ (\alpha+\delta) t -\frac 12 I  \right] O(t) \right\ra_{\text{eq}} \cr
&=& \beta^2 \left[(\alpha+\delta) t \la J_{\text{in}} \, O(t)\ra_{\text{eq}}  - \frac 12 \la J_{\text{in}}\, I \, O(t)\ra_{\text{eq}} \right].\label{eq:pert1}
\eea

\begin{figure}[t]
 \centering
 \includegraphics[width=10 cm]{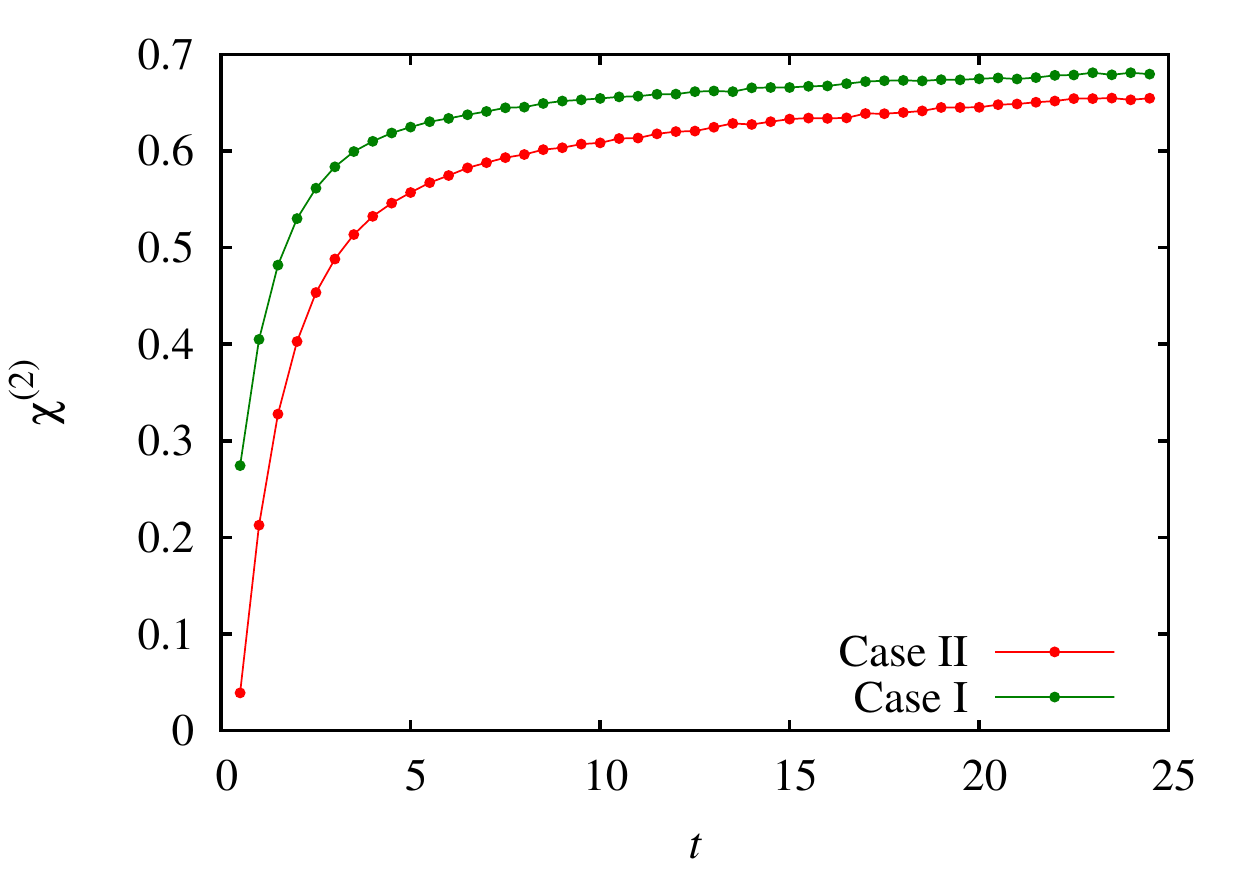}
 \caption{The second order response \eqref{eq:pert1} and \eqref{eq:pert2} for two kinetically different but thermodynamically equal perturbations in the open zero range process. The observable considered is the particle number $n_1(t)$ at the first site at time $t.$ The injection and removal rates at the boundary are $\alpha=2.5,\gamma=5.0$ and $\delta=4.0, \kappa=8.0$ for the unperturbed system and $\beta=1$. In the bulk the particles hop symmetrically with unit rate. The lattice size is $L=10$ in both cases. Not shown is the linear response which is identical for the two cases.}
 \label{fig:zrp_compare}
\end{figure}

{\bf Case II.} Next we consider a different perturbation where both the entry and the exit rates are modified $\alpha \to \tilde \alpha = \alpha e^{\beta \varepsilon/2}/(1+\beta \varepsilon), \gamma \to \tilde \gamma = \gamma e^{-\beta \varepsilon/2}/(1+\beta \varepsilon)$ and $\delta \to \tilde \delta = \delta e^{\beta \varepsilon/2}/(1+\beta \varepsilon), \kappa \to \tilde \kappa = \kappa e^{-\beta \varepsilon/2}/(1+\beta \varepsilon)$.  It gives rise to the same shift in the chemical potential $\tilde{\mu}=\mu + \varepsilon$ as before, and the entropic contribution $S_0'(\omega ) $ remains the same. But the frenetic part has a completely different form now,
\bea
D_0^\prime(\omega) = \beta  \left[I  - \frac 12 (\alpha+\delta) t -\frac 32 \int_0^t \id s~[\gamma u(n_1(s))+ \kappa u(n_L(s))]\right]
\eea
The third term measures the outflux from the system to the environment during time $t.$ Thus the second order response for the observable $O(t)$ takes the form,
\bea
\la S_0'(\omega ) D_0'(\omega ) O(t) \ra &=& \beta^2 \left[\la J_{\text{in}}I;O \ra_{\text{eq}} - \frac 12 (\alpha+\delta) t \la J_{\text{in}} ; O(t)  \ra_{\text{eq}} \right. \cr
&& -\frac 32  \left. \int_0^t \id s~ \la [\gamma u(n_1(s))+\kappa u(n_L(s))]\,J_{\text{in}}; O(t)\ra_{\text{eq}}\right] \label{eq:pert2}
\eea
Equations \eqref{eq:pert1} and  \eqref{eq:pert2} give the response formula for two perturbations which are equivalent thermodynamically (a shift in the chemical potential of the environment). One can design a variety of perturbations with the same property but very different in  kinetic factors. They would result in the same linear response, but when it comes to  second order the kinetic aspects like the exact details  of the coupling to the reservoirs show up.

This is illustrated in Fig.~\ref{fig:zrp_compare} where plots of the second order response as obtained from numerical simulations using Eqs. \eqref{eq:pert1} and  \eqref{eq:pert2} are shown, with
\bea
\chi^{(2)} &:=& \frac 1{\varepsilon^2} \left[\langle n_1(t)\rangle - \langle n_1(t) \rangle_{\text{eq}} - \varepsilon \beta \la n_1(t) J_\text{in} \ra_{\text{eq}} \right], \quad \varepsilon\downarrow 0 
\eea
where  $O(t)= n_1(t)$ is the particle number at the left boundary site. Its expected value 
 is considerably affected by the kinetic details of the perturbation, as seen by the two distinct curves in Fig. \ref{fig:zrp_compare}, but \underline{only} from second order response on. There are of course  to all orders no differences between cases I and II in the $t\downarrow 0$ (initial condition) and the $t\uparrow \infty$ (new equilibrium) limits.

%


\subsection{Second order dielectric response}\label{diel}
Static external fields allow for inhomogeneous equilibrium systems, e.g. with gradients in density or  energy. In such cases, detailed balance is maintained and currents caused by e.g. gravity or electrostatics cancel with the diffusive currents.  An important example is the phenomenon of polarization, where opposite electric charges
are being separated by an external field $\varepsilon$.  If we however impose that field in a time-dependent way, the system gets out-of-equilibrium showing time-dependent charge displacement and we can measure the response via the frequency dependence in the dielectric permittivity.  Second order effects have been investigated for specific contexts in Refs.~\cite{pine,mik,mik2} where one uses the diagonalization of the unperturbed Hamiltonian.  Here we investigate how changes in dynamical activity influence the second order response.

\begin{figure}[ht]
 \centering
 \includegraphics[width=8 cm,bb=0 0 294 314]{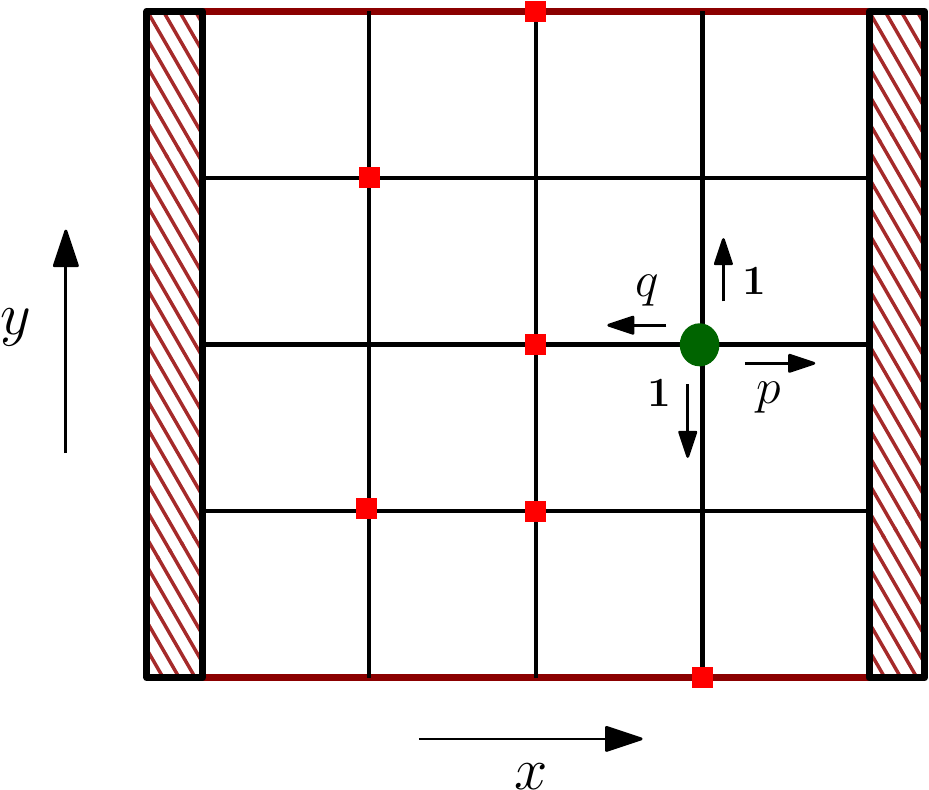}
 \caption{Lorentz model for polarization. Red squares denote the obstacles and the green circular disc denotes the biased random walker.}
 \label{fig:lorentz}
\end{figure}

Take a particle system in a box in equilibrium.  Particles are either positively or negatively charged but we will forget about the internal field, thus neglecting mutual particle interactions.  We apply a time-dependent but spatially uniform electric field $\varepsilon\,h(s)$ for $s>0$ when starting from equilibrium.  The polarization ${\cal P}(t)$ measures the local gradient in charge density at time $t$. To be specific  we choose a particular set-up reminiscent of the Lorentz model for free charge carriers in a disordered lattice. Independent walkers (dilute electron gas) hop in continuous time to neighboring sites on a lattice square with $L \times L$ sites which is closed in the direction of the applied field (say $x$-direction) and periodic in the orthogonal ($y$) direction.  The field leads to a biased random walk in the $x$-direction  --- at time $s>0$ the particle moves to the right (left) neighboring site with rate $p_s(q_s)$ while motion in the $y$-direction is unbiased. The system is disordered 
by randomly removing lattice sites, 
thus introducing obstacles for the walkers.
A particle can move only when the target site does not have an obstacle on it. Fig. \ref{fig:lorentz} schematically describes the system. We choose the perturbed rates for horizontal jumps at times $s>0$ as 
\bea\label{ps}
p_s= \frac{e^{\beta \varepsilon h_s/2}} {e^{\beta \varepsilon h_s/2} +e^{-\beta \varepsilon h_s/2}}; \quad q_s= \frac{e^{-\beta \varepsilon h_s/2}} {e^{\beta \varepsilon h_s/2} +e^{-\beta \varepsilon h_s/2}} 
\eea
which satisfies the demand of local detailed balance $p_s/q_s = e^{\beta \varepsilon h_s}.$ 


For general protocol $h_s$ we need the extension in Appendix \ref{genca} of the formalism in Section \ref{path}.  We then find the change in entropy from \eqref{soa}, as the  sum over jump times $s_i$
\bea\label{39}
S_0'(\omega) = \beta \sum_{s_i} h_{s_i}(x_{s_i} - x_{s_i^-}) =\beta \int_0^t\id s \,h_s\,j_s^x(\id \omega)
\eea
where $x_s$ denotes the $x$-coordinate of the particle at time $s.$ (Note: $x$ is not the configuration here.) The current $j_s^x(\id \omega) = \pm 1$ when a right/left horizontal jump occurs at time $s$ and is zero otherwise. The change in the dynamical activity follows from \eqref{eq:Dprimea} where $\phi_s\equiv 0,$
\bea\label{40}
D_0'(\omega) = \frac{\beta}{4} \int_0^t \id s~ h_s (\eta_{x_s-1} - \eta_{x_s+1}) 
\eea
Here $\eta$ denotes the presence of the obstacles: $\eta_i=1,0$ depending on whether there is an obstacle at the $i^{th}$ site or not ($y$ coordinate is suppressed). We can see it as a directed collision frequency $C^x(s) := \eta_{x_s-1} - \eta_{x_s+1}$.
Again $x$ is the horizontal coordinate. The jumps along the $y$-direction do not explicitly contribute in the response. 

\begin{figure}[t]
 \centering
 \includegraphics[width=8 cm]{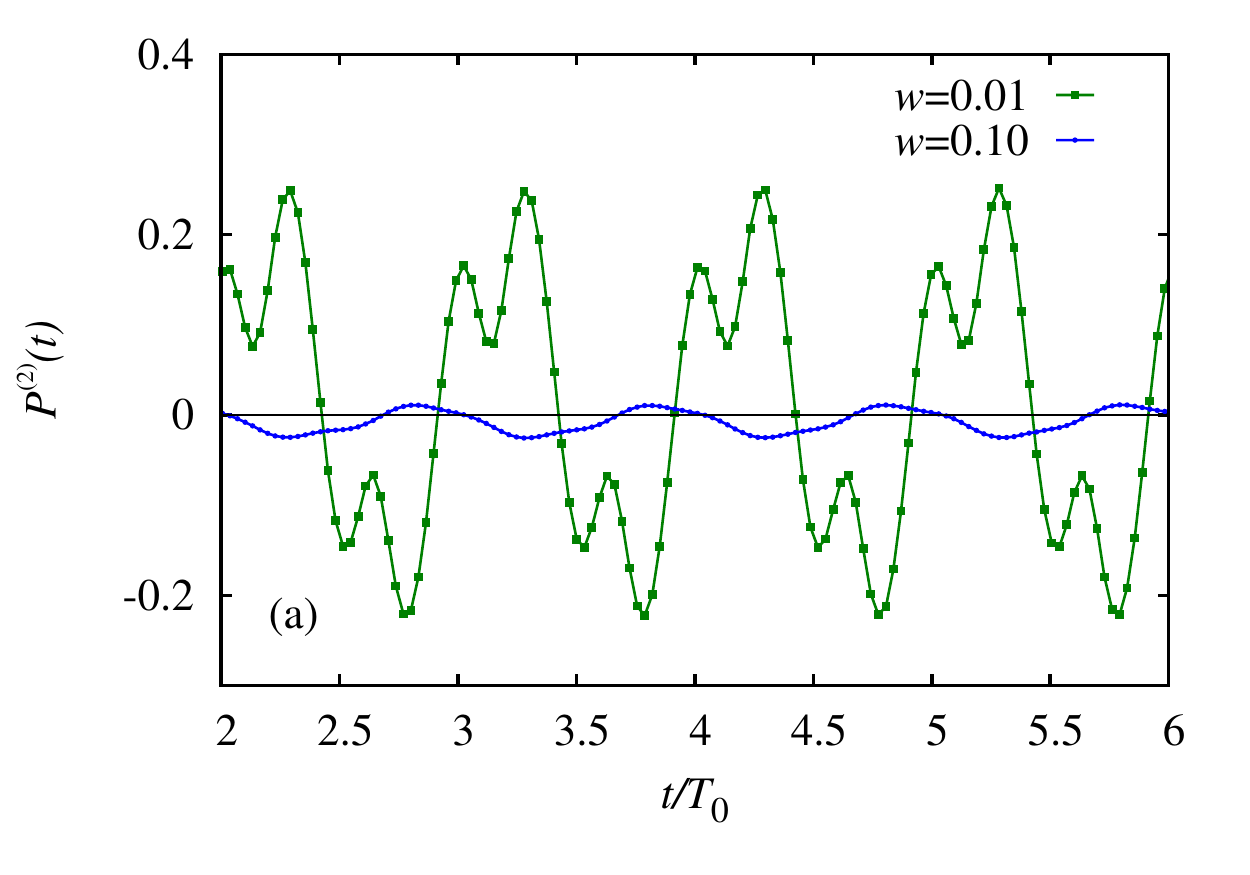}\includegraphics[width=8 cm]{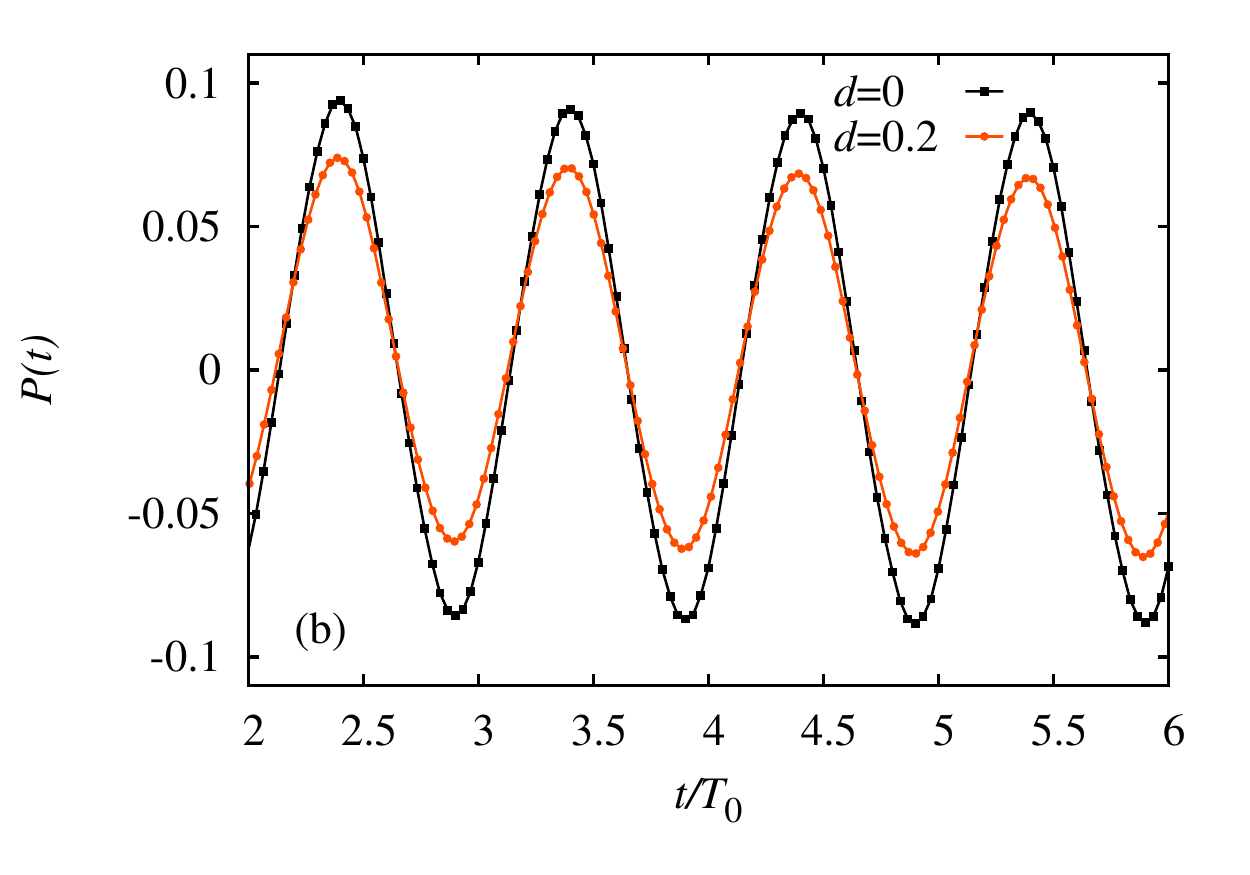}
 \caption{(a)  The second order correction  $P^{(2)}$ in the density gradient {\it versus} time $t/T_0$ for two different frequencies $w$. Here $T_0=2 \pi/ w$ is the time period.  The obstacle density is $d=0.2$. In case of no obstacles, there is no second order response. Note also the typical emergence of higher harmonics.  For large frequency the second-oder contribution vanishes.  (b) Comparison of the density gradient $P$ in the presence (circles) and absence (squares) of obstacles. Here frequency $w=0.1$ and the perturbation strength $\varepsilon=0.5.$ }
 \label{fig3}
\end{figure}

  The dielectric constant is given by the response of the polarization as measured e.g. in the density gradient.  The linear response is made from the equilibrium correlation function between the polarization and the dissipative currents $j_s^x(\id \omega)$ in \eqref{39} created by the field.  In other words, that correlation function involves observables that are not explicitly related to changes in dynamical activity or to the presence of obstacles.  (Of course the equilibrium statistics itself does depend on the density of obstacles.) On the other hand, the second order explicitly takes into account how  the dynamical activity is influenced by the field.   For the response in observable $O$ there appears in second order the correlation $\langle O(t) j_s^x(\id \omega)\,C^x_{s'}\rangle$ where the collision frequency defined below Eq. \eqref{40} enters explicitly.  In Fig.~\ref{fig3}(a) we have $h_s\propto \sin ws$ in the biased rates \eqref{ps} for the random walkers, showing the second order $P^{(2)}$ in 
the polarization, i.e., in the difference between particle densities at the left and right edge in Fig.~\ref{fig:lorentz}.
   
 We may also note that the dielectric response is decreased by the disorder; Fig~\ref{fig3}(b) shows a comparison of the polarizations in presence and absence of obstacles.  The main underlying reason here is that the particles get more easily trapped between obstacle configurations, which is a kinetic effect. The consequence of this is that both the current and dynamical activity decrease in the presence of disorder. The same mechanism is at work for producing negative differential conductivity in such systems, \cite{fren}.  The point is that an increase in the external field can increase the trapping of particles. \\
  We further illustrate that effect in Appendix \ref{compu} via a hydrodynamic description in terms of a one-dimensional charge density $\rho(x,t), x\in [0,1]$.

\subsection{Second order mobility in narrow channels}
The Sutherland--Einstein relation gives the linear response relation between mobility and diffusion.  We investigate here the second order correction for particles in a one--dimensional periodic potential which e.g. describes the motion of particles in channels. We take the evolution equation with friction $\gamma$ at temperature $T$,
\[
\dot x_t = v_t,\quad \dot v_t = -\gamma v_t + \varepsilon - U'(x_t) + \sqrt{2\gamma T}\,\xi_t
\]
for position $x_t\in \R$ with velocity $v_t$ and mass $m=1$.  The potential $U$ is periodic and the perturbation $\varepsilon$ is switched on at $t=0$.  For this dynamics the Sutherland-Einstein relation $M=\hat{\cal D}/T$ is verified for the equilibrium values ($\varepsilon=0$) of the resulting, time dependent diffusion coefficient ${\hat{\cal D}}$ and mobility $M$,
\[
\hat{\cal D}(t) = \frac 1{2t}\langle (x_t-x_0)^2\rangle_{\text{eq}},\quad M(t) = \frac{\partial}{\partial \varepsilon}\frac 1{t}\langle x_t-x_0\rangle_{\varepsilon=0}.
\]
Note that $\hat{\cal D}(t)$ is not to be confused with the ``bare'' $\cal D$ in Eq.~\eqref{overd} or Eq.~\eqref{eq:be} below. Furthermore, slightly different definitions for ${\hat{\cal D}}(t)$ and $M(t)$ than the ones above have been used in the literature, which however plays no crucial role in our discussion. The displacement $\langle x_t-x_0\rangle$ is odd in $\varepsilon$ when the landscape is spatially reflection--invariant.  In order to have a non-trivial second order, we need a potential $U$ that breaks that reflection--invariance.  Furthermore, there is no confined stationary condition here, as for large times the particle diffuses uniformly on $\R$. We can nevertheless continue to apply the formalism of Section \ref{path}.\\
The action is 
\[
{\cal A}=-\frac {\varepsilon}{2\gamma\,T}\left\{ v_t-v_0 + \gamma\int_0^t v_s\id s + \int_0^t U'(x_s)\,\id s \right\}
\]
to first order in $\varepsilon$; see also Ref.~\cite{sog} for the case of visco-elastic media.  In other words,
\[
D_0'= -\frac {1}{2\gamma\,T}\{v_t-v_0 + \int_0^t U'(x_s)\,\id s\},\quad S_0' = \frac {1}{T}\int_0^t v_s\id s = \frac{x_t-x_0}{T}
\]
which together determine the second order response.  For the mobility $M(t)$ the observable is $j = (x_t - x_0)/t$ in \eqref{gk}, and we find
\[
\langle x_t-x_0\rangle = \frac{\varepsilon}{2T} \langle (x_t-x_0)^2\rangle_{\text{eq}} +  \frac{\varepsilon^2}{4\gamma T^2} \left\langle (x_t-x_0)^2\,\{v_t-v_0 + \int_0^t U'(x_s)\,\id s\}\right\rangle_{\text{eq}}. 
\]
The second order correction to the Sutherland--Einstein relation around equilibrium carries information about the friction constant $\gamma$ and about the correlation between square displacement (which can be called a local diffusion constant) and the periodic force:
\[
\frac 1{t\varepsilon}\langle x_t-x_0\rangle  = \frac{1}{T} \,\hat{\cal D}(t) +  \frac{\varepsilon}{2\gamma T^2} \left\langle \frac{(x_t-x_0)^2}{2t}\,\left \{v_t-v_0 + \int_0^t U'(x_s)\,\id s \right \} \right\rangle_{\text{eq}} .
\]
Note that $v_t-v_0 + \int_0^t U'(x_s)\,\id s$ exactly measures the deviation from the mechanical second  Newton's law as visible in the trajectories of the Langevin particle. (We have taken the mass $m$=1). See also Refs.~\cite{kru1,bli,proc} for nonequilibrium modifications of the Sutherland--Einstein relation.\\

For an overdamped version of the above where the correlation with the local diffusion constant gets even more visible, we can study the dynamics \eqref{overd} with time-independent perturbing potential $V(x) =x$ switched on at $t=0$, and we take potential $U$ periodic (and not confining). Motivated from the above, we study the case of a  spatially varying diffusion coefficient ${\cal D}(x)$, which can e.g. emerge through hydrodynamic interactions with boundaries.  Here, we approximatively interpret it as to arising from a local temperature; we call ${\cal D}'$ the Soret force which in equilibrium is balanced by the ``entropic'' diffusion force. (Note that there is no equilibrium for temperature inhomogeneities, and this notion identifies the approximative nature of our interpretation: It assumes a local equilibrium condition.)\\
We calculate again \eqref{sd} and \eqref{16} for $V(x)=x$.  The entropic and frenetic components are
\beq
S'_0(\omega) = \beta(x_t-x_0),\quad D'_0(\omega) = {\beta\over 2}\,\int_0^t\id s\, [-\chi U' + {\cal D}'](x_s)  .\label{eq:be}
\eeq
Even for $U'\equiv 0$ we can get a second order contribution to the mobility when the ``temperature'' gradient ${\cal D}'$ is not reflection invariant:
\[
\langle x_t-x_0\rangle = \frac{\beta\varepsilon}{2}\langle (x_t-x_0)^2\rangle_{\text{eq}} +  \frac{\beta^2\varepsilon^2}{4}\int_0^t\id s\, \langle (x_t-x_0)^2\,{\cal D}'(x_s) \rangle_{\text{eq}}.
\]

We see here that the Soret force couples with the square displacement in the equilibrium expectation to determine the displacement caused by the field in second order; see also \cite{lib}.
There are however more obvious possibilities of segregation of driven particles based  on their (excess) dynamical activity.  A good example can be found in \cite{beh,clem}.  There, gold-capped colloidal spheres are suspended in a critical binary liquid mixture and perform active
Brownian motion when illuminated by light.  One can change the functionalization of the gold cap, meaning to change their distance from equilibrium. The swimmers under larger driving are further away from equilibrium and therefore, consistent with the results of the present paper, undergo the influence of the change in dynamical activity.  In that regime in the presence of obstacles and because of the frenetic contribution in the response which reduces the effective escape rate, the current drops and the swimmers are deflected orthogonally from their original direction.  They leave the system before reaching the end.  On the other hand, the less functional swimmers are still in the linear regime where the coupling with dynamical activity is absent and continue their motion reaching the end.

\section{Remarks}
After finishing these specific cases and models, we end with a couple of general remarks that continue from Section \ref{path}.
\begin{enumerate}
\item
There are obviously cases where the second order response is the leading response, in the sense that the linear response equals zero.  That depends on perturbation, observable and symmetries.  In that case, the kinetic effects become directly visible as most significant in the response.  A well known class of examples are unloaded ratchet systems \cite{woj}. 
In so far as these ratchet systems are relevant in life processes (e.g. as models of molecular motors, \cite{mol}), the notion of dynamical activity and hence frenetic aspects will also be  so.\\

\item
There are general ways to connect the various orders of nonlinear response, see e.g. \cite{and}.  For example, a second order response can be expressed as the linear response to a suitable observable.  Section 6 in Ref.~\cite{col} gives some explicit calculation.  Looking for example at \eqref{gk} and applying it first to $J\rightarrow JD_0'$:
\[
\frac{\partial}{\partial \varepsilon}
\langle J\, D'_0\rangle_{|_{\varepsilon=0}} = \frac 12\langle D'_0\,S'_0\, J\rangle_{\text{eq}} = - \frac 12 \frac{\partial^2}{\partial \varepsilon^2}
\langle J\rangle_{|_{\varepsilon=0}}
\]
meaning that the linear response to $ J\, D'_0$ is proportional to the second order response to $J$. That implies that the second order response takes into account how the dynamical activity changes under the perturbation. Note however that it is experimentally often more convenient to measure second order responses around equilibrium than to measure linear responses around nonequilibrium.

\item Note that $D'_0$ and $S'_0$ are not entirely independent as by construction $\langle {\cal A}'_0\,
 O(x_0)\rangle_{\text{eq}}=0$ (for any observable $O$ in \eqref{2n} that depends on the state $x_0$ at time zero only).  That implies 
 that in the equilibrium process conditioned on the beginning,
 \begin{equation}\label{canc}
2 \langle  D'_0 |x_0=x \rangle_{\text{eq}} = 
  \langle  S'_0 |x_0=x \rangle_{\text{eq}}
  \end{equation}
for all initial states $x$.  For overdamped diffusions the relation \eqref{canc} is clearly visible when comparing \eqref{sd} with \eqref{16}.
Note that \eqref{canc} implies $\langle  D'_0 \rangle_{\text{eq}} = 0$ so that from \eqref{2n8}, 
to first order always
\begin{equation}\label{2n88}
\langle D'_0\rangle =  - \varepsilon \,\langle (D'_0)^2
\rangle_{\text{eq}} \leq 0.
\end{equation} 

\end{enumerate}

\section{Conclusions}
We presented a theoretical framework that conveniently computes the second order response via path integrals, thereby providing explicit examples of how experimentally measured quantities in equilibrium determine the second order response around equilibrium. While the general statement hence is that second order response can be derived in terms of equilibrium correlations, just as in linear response, we point to distinct differences with the latter: Certain kinetic effects appear explicitly only starting from the second order. We have called them frenetic aspects as the dynamical activity is the unifying observable that enters in correlation with entropy fluxes and observables. The theory can prove useful to understand different responses resulting from kinetic effects or, inversely, to separate different types of particles according to their second order response.  The very fact that the (time-symmetric) dynamical activity can change under (nonequilibrium) perturbations is important and manifest in many examples \cite{fren,gen,gene}, and can now be the basis for further practical purposes.\\

We furthermore discussed the connection between second order responses around equilibrium and linear responses around nonequilibrium, thus complementing recent work on nonequilibrium linear response \cite{sei,lips,fdr,kru2}. As another relation with existing work, we note that
second or higher order response cannot be directly derived in terms of equilibrium correlations from the fluctuation symmetry \cite{Ga} in the distribution of the entropy flux, nor does its physical meaning emerge clearly from a Dyson expansion.   Here we have given an alternative approach with the advantage of bringing to the foreground the role of excess dynamical activity.

\vspace{1cm}
\noindent {\bf Acknowledgment}\hspace{1cm}
This work was financially supported by the Belgian Interuniversity
Attraction Pole P07/18 (Dygest) and by the DFG grant No. KR 3844/2-1.
\newpage

\appendix

\section{Time-dependent perturbations}\label{genca}
We extend here the analysis of Section \ref{path} to include time-dependent perturbations.  More in particular, we will show the general validity of \eqref{kubo2}, as we have used it in Section \ref{diel}.  We do not know whether also the second order Green-Kubo formula \eqref{gk} holds true for general time-dependent protocols.\\
To begin and to illustrate the situation for the simplest case let us revisit Section \ref{od} with time--dependent potential perturbations.  We now have \eqref{od} changed into
\begin{equation}
\label{overda}
\dot x_s = \chi(x_s)\,f_s(x_s) + \nabla {\cal D}(x_s) +\sqrt{2{\cal D}(x_s)}\,\xi_s 
\end{equation}
 with $f_s(x) = -\nabla U(x) + \varepsilon h_s\,\nabla V(x)$ where we add the time--dependent amplitude $\varepsilon h_s$ for positive times $s>0$.\\  
Path-integral techniques \cite{fdr1} deliver the action ${\cal A}$ as needed for the expansion e.g. in \eqref{2n}, but the time-reversal must now also include the time-reversal of the protocol $h_s \rightarrow h_{t-s}$ for the decomposition ${\cal A} = D - \frac 1{2} S$ to have physical meaning.  Indeed, only if we include that protocol-reversal in the time-reversal do we find the physically correct entropic component
\[
S_\varepsilon(\omega) = \varepsilon\,\beta\,\int_0^t h_s\,\nabla V(x_s) \circ \id x_s
\]
in terms of a Stratonovich integral, so that by the Fundamental Theorem of stochastic calculus
\begin{equation}\label{sda}
S'_0(\omega) =\beta[h_tV(x_t)-h_0V(x_0) - \int_0^t\id s \,\dot{h}_s\,V(x_s)].
\end{equation}
Comparing that with the First Law, $S'_0(\omega)/\beta$ must be the heat so that we indeed recognize in $S'_0(\omega)$  the excess entropy flux (per $k_B$) to the heat bath.\\
The frenetic component equals
\begin{equation}\label{16a}
D'_0(\omega) = {\beta\over 2}\,\int_0^t\id s\, h_s\, [-\chi\nabla U\cdot \nabla V + \nabla\cdot ({\cal D}\nabla V)](x_s) = {\beta\over 2 }\,\int_0^t\id s\, h_s\, LV(x_s).
\end{equation}
As a consequence the above \eqref{sda} and \eqref{16a} are in general no longer antisymmetric, respectively symmetric, under the (naked) time-reversal $\theta$ of Section \ref{path}, which implies that the derivations there must be modified for e.g.~\eqref{kubo2} still to be correct also with \eqref{sda}--\eqref{16a}.\\ 

Similarly, for the Markov jump case we consider
\begin{equation}
k_{s}(x,y)= k_{\text{eq}}(x,y)\,[1+\varepsilon\phi_s(x,y)+ \frac{\varepsilon}{2}a_s(x,y) + O(\varepsilon^2)]\label{paa}
\end{equation}
with time-dependent $a_s(x,y)$ and $\phi_s(x,y), s>0$.
The entropic component obtained by reversing time also in the protocol now equals
\begin{equation}\label{soa}
S'_0(\omega) = \sum_{s_i} a_{s_i}(x_{s_{i^-}},x_{s_i})
\end{equation}
 and for a  perturbation via the addition of a potential $V$ with time-dependent amplitude $h_s$, we get as in \eqref{sda}
\begin{equation}\label{ssa}
a_s(x,y) = \beta
h_s \,[V(y) - V(x)] \;\implies\; S'_0(\omega) =\beta[h_tV(x_t)-h_0V(x_0) - \int_0^t\id s \,\dot{h}_s\,V(x_s)]
\end{equation}
Similarly, for the parameterization \eqref{paa} we find the frenetic component
\bea
D'_0(\omega) = -\sum_{s_i} \phi_{s_i}(x_{s_{i^-}},x_{s_i}) + \int_0^t\id s \sum_y k_{\text{eq}}(x_s,y)\,[\phi_s(x_s,y) + \frac 1{2} a_s(x_s,y)] \label{eq:Dprimea}
\eea
For these jump processes the equality \eqref{canc} is realized by the identities
\begin{eqnarray}
\langle\sum_{s_i} \phi_{s_i}(x_{s_{i^-}},x_{s_i})|x_0=x\rangle_{\text{eq}} &=& \int_0^t\id s \sum_y \langle k_{\text{eq}}(x_s,y)\,\phi_s(x_s,y)|x_0=x\rangle_{\text{eq}}\nonumber\\
\langle \sum_{s_i} a_{s_i}(x_{s_{i^-}},x_{s_i})|x_0=x\rangle_{\text{eq}} &=& \int_0^t\id s \sum_y \langle k_{\text{eq}}(x_s,y)\, a_s(x_s,y)|x_0=x\rangle_{\text{eq}}.
\end{eqnarray}

We now re-derive the second order Kubo formula \eqref{kubo2} along the lines of Section \ref{path} but with \eqref{sda}--\eqref{16a} or \eqref{soa}--\eqref{eq:Dprimea}. Remember that the action ${\cal A}_\varepsilon = D- S/2$ enters the expansion \eqref{2n}.  We start with the linear response where for \eqref{kubo2}  we need to show $\langle {\cal A}'_0(\omega)\, O(x_t) 
\rangle_{\text{eq}}  = - \langle S'_0(\omega)\, O(x_t) 
\rangle_{\text{eq}}$, see \eqref{2n}.  It suffices therefore that $\langle D'_0\, O(x_t)\rangle_{\text{eq}} = -\frac 1{2} \langle S'_0\, O(x_t)\rangle_{\text{eq}}$.  That follows from
\begin{eqnarray}\label{loa}
\langle D'_0(\omega)\, O(x_t)\rangle_{\text{eq}} &=& \langle {\tilde D}'_0(\omega)\,O(x_0)\rangle_{\text{eq}} \nonumber\\
&=& \frac 1{2}\langle {\tilde S}'_0(\omega)\,O(x_0)\rangle_{\text{eq}} =
-\frac 1{2} \langle {S}'_0(\omega)\,O(x_t)\rangle_{\text{eq}}
\end{eqnarray}
where the first and last equality make use of the tilde-observables obtained from the original ones by time-reversing the protocol; for example, $\tilde{S}(\theta\omega) = -S(\omega), \tilde{D}(\theta\omega) = D(\omega)$. The middle equality in \eqref{loa} is the identity \eqref{canc} for time-reversed protocol.\\
For the second order response we  use the second order version of \eqref{canc} saying that for all protocols $\langle \left[{\cal A}''_0(\omega)-{\cal A}'_0(\omega)^2\right]\, O(x_0)\rangle_{\text{eq}}=0$
which implies that
\begin{equation}\label{2v}
\left\langle \left[{\tilde D}^{''}_0 - ({\tilde D}'_0)^2 -  \frac 1{4}({\tilde S}'_0)^2\right]\,O(x_0) \right \rangle_{\text{eq}} = -
\langle {\tilde D}'_0\,{\tilde S}'_0\,O(x_0)\rangle_{\text{eq}}
\end{equation}
again in terms of the tilde-observables.  The identity \eqref{2v} follows from \eqref{2n} by taking $O(\omega)=O(x_0)$ to depend on the initial time only. On the other hand, the left-hand  side of \eqref{2v} equals
\[
\left \langle \left[D^{''}_0 - (D'_0)^2 -  \frac 1{4}(S'_0)^2\right]\,O(x_t)\right\rangle_{\text{eq}} =
\left\langle \left[\tilde{D}^{''}_0 - (\tilde{D}'_0)^2 -  \frac 1{4}(\tilde{S}'_0)^2\right]\,O(x_0)\right\rangle_{\text{eq}}
\]
while its right-hand side equals
\[
-\langle {\tilde D}'_0\,{\tilde S}'_0\,O(x_0)\rangle_{\text{eq}} =  \langle D'_0\,S'_0\,O(x_t)\rangle_{\text{eq}}.
\]
Hence, combining the above
\begin{eqnarray}
\left \langle \left[{\cal A}''_0(\omega)-{\cal A}'_0(\omega)^2\right]\, O(x_t)\right \rangle_{\text{eq}} &=&
\left \langle \left[{D}^{''}_0 - ({D}'_0)^2 -  \frac 1{4}({S}'_0)^2\right]\,O(x_t) \right \rangle_{\text{eq}} + 
 \langle {D}'_0\,{S}'_0\,O(x_t) \rangle_{\text{eq}} \nonumber\\
&=& \left \langle \left[{\tilde D}^{''}_0 - ({\tilde D}'_0)^2 -  \frac 1{4}({\tilde S}'_0)^2\right]\,O(x_0) \right \rangle_{\text{eq}}  + 
\langle {D}'_0\,{S}'_0\,O(x_t)\rangle_{\text{eq}}\nonumber\\
&=& -
\langle {\tilde D}'_0\,{\tilde S}'_0\,O(x_0)\rangle_{\text{eq}} + 
\langle {D}'_0\,{S}'_0\,O(x_t)\rangle_{\text{eq}}\nonumber\\
&=& 
2\langle {D}'_0\,{S}'_0\,O(x_t)\rangle_{\text{eq}}
\end{eqnarray}
which proves \eqref{kubo2} for general perturbation protocols.  It is an open question whether \eqref{gk} is equally valid beyond linear response for arbitrary time-dependent perturbations.

\section{Response formalism}\label{compu}

The point of the paper is not merely to compute second order responses but to provide more physical or systematic insight. Yet for the sake of completeness, we bring here the more standard computational scheme, if only for comparison with the formalism of the present paper.

\subsection{Perturbation analysis}

The analytic approach to compute nonlinear response for a Markov system  is to look at the Taylor expansion for the probability density $\rho(x,t).$  Let us consider an overdamped diffusion process described by the Smoluchowksi equation 
\[
\frac{\id \rho}{\id t} = L^\dagger \rho
\]
in terms of the forward generator $L^\dagger$.
Writing there both the density and the generator as a series in the perturbation parameter $\varepsilon$,
$\rho=\rho_0+\varepsilon\rho_1+\varepsilon^2\rho_2~..., \;
L^\dagger=L^\dagger_0+\varepsilon L^\dagger_1$,
we obtain
\begin{align}
\frac{\id \rho_0}{\id t} &= L^\dagger_0 \rho_0,\quad
\frac{\id \rho_1}{\id t} = L^\dagger_0 \rho_1+L^\dagger_1 \rho_0\nonumber\\
\frac{\id \rho_2}{\id t} &= L^\dagger_0 \rho_2+L^\dagger_1 \rho_1
\end{align}
which can be solved order by order.  Equivalently, we can start from the formal solution to the Smoluchowski equation, starting from the  equilibrium $\rho_{\text{eq}}$,
\begin{equation}
\rho=\exp_+\biggl[\int_{0}^{t}L^\dagger\biggr] \rho_{\text{eq}}
\end{equation}
where $\exp_+$ stands for the time-ordered exponential. Expansion yields the usual terms in the Dyson series 
\begin{align}
\rho_1&=\int_{0}^{t}\id t_1 \,e^{(t-t_1)L_0^\dagger}L_1^\dagger(t_1) e^{t_1\,L_0^\dagger}\rho_{\text{eq}}\nonumber\\
\rho_2&=\int_{0}^{t}\id t_2\int_0^{t_2}\id t_1 \,e^{(t-t_2)L_0^\dagger}L_1^\dagger(t_2) e^{(t_2-t_1)L_0^\dagger}L_1^\dagger(t_1) e^{t_1\,L_0^\dagger}\rho_{\text{eq}}.
\end{align}
After an infinite time, $t \uparrow \infty$, we are looking at the new stationary regime and we write the different orders of the stationary probability density as
\begin{align}
\rho_0&=\rho_\text{eq}, \quad
\frac{\rho_1}{\rho_\text{eq}}=-\frac{1}{L_0}\Bigl\{\frac{1}{\rho_\text{eq}}L_1^\dagger \rho_\text{eq}\Bigr\}\nonumber\\
\frac{\rho_2}{\rho_\text{eq}}&=\frac{1}{L_0}\Bigl\{\frac{1}{\rho_\text{eq}}L_1^\dagger \rho_\text{eq}\frac{1}{L_0}\Bigl\{\frac{1}{\rho_\text{eq}}L_1^\dagger \rho_\text{eq}\Bigr\}\Bigr\}
\end{align}
where $\{\centerdot\}$ represents the fluctuating part of a function (i.e., that function minus its average), and $\frac{1}{L_0}\{\centerdot\}$ is the so-called pseudo-inverse of the backward generator $L_0$. Obtaining these expressions has relied solely on using the detailed balance condition $L^\dagger_0 \rho_\text{eq}=\rho_\text{eq} L_0$. The first order gives the so-called McLennan distribution \cite{McL}. The appearance of the pseudo-inverse in these expressions makes it more difficult to use them systematically or explicitly.\\

We now look at the more specific case of a  potential perturbation of an overdamped diffusion. The forward Smoluchowski generator changes to $L^\dagger(s) = L^\dagger_0 + \varepsilon h_s\,L^\dagger_1$, with $L^\dagger_1 \rho = -\nabla (\nabla V \rho)$.\\
To calculate the response in some observable it is more convenient to work with the backward generator $L$, which is the adjoint of $L^\dagger$. The expectation of a state observable $O$ at time $t$ in the perturbed system is given by 
\bea
\la O(t) \ra = \int \id x~ \rho_\text{eq}(x)~ \exp_{-}^{\int_0^t \id s L(s)} O(x). \label{eq:time}
\eea
Here the system starts at $t=0$ in the equilibrium density $\rho_{\text{eq}}$, and $\exp_{-}$ is the anti-ordered exponential. The perturbed backward generator is $L(s) = L_0 + \varepsilon h_s L_1$ with 
\bea
L_1 =  \nabla V \nabla = -\frac 12 [V L_0 - L_0 V +(L_0V)]. 
\eea
Let us write
\[
\la O(t) \ra = \la O(t) \ra_\text{eq} + \varepsilon~ \chi^{(1)} + \varepsilon^2 ~ \chi^{(2)}+...
\]
From the above, and using the fact that $\rho_\text{eq}L_1=-\rho_\text{eq}VL_0$, we get
first the Kubo formula (we assume $\beta=1$ for the sake of simplicity)
\[
\chi^{(1)} = \int_0^t \id s ~ h_s \frac{\id}{\id s} \la V(s) O(t) \ra_\text{eq}
\]
and next, for second order,
\bea
\chi^{(2)} &=& \frac 12 \int \id x ~\rho_{\text{eq}}(x) \int_0^t \id t_1 h_{t_1}  \int_0^{t_1} \id t_2 h_{t_2} V L_0 e^{(t_1 -t_2)L_0} [V L_0 - L_0 V + (L_0V)]e^{(t-t_1)L_0}O(x) \cr
&=& -\frac 12 \left[  \int_0^t \id t_1~ h_{t_1}  \int_0^{t_1} \id t_2~ h_{t_2} \frac{\id}{\id t}  \frac{\id}{\id t_2} \la V(t_2) V(t_1) O(t) \ra_\text{eq} \right. \cr
&&~~~ + \int_0^t \id t_1~ h_{t_1}  \int_0^{t_1} \id t_2~ h_{t_2} \frac{\id^2}{\id t_2^2} \la V(t_2) V(t_1) O(t) \ra_\text{eq}  \cr
&&~~~  \left. + \int_0^t \id t_1~ h_{t_1}  \int_0^{t_1} \id t_2~ h_{t_2}  \frac{\id}{\id t_2} \la V(t_2) (L_0V)(t_1) O(t)\ra_\text{eq} \right].
\eea
The first two terms can be combined into
\bea
-\frac 12 \int_0^t \id t_1~ h_{t_1} \int_{t_1}^t \id t_2 h_{t_2}  \frac{\id}{\id t} \frac{\id}{\id t_1} \la V(t_1) V(t_2) O(t) \ra_\text{eq}
\eea
and adding the third term yields
\bea
\chi^{(2)} = -\frac 12 \int_0^t \id t_1~ h_{t_1} \int_0^t \id t_2~ h_{t_2} \frac{\id}{\id t_2} \la V(t_2) (L_0V)(t_1) O(t) \ra_\text{eq}.
\eea
This expression is identical to the second order response formula \eqref{kubo2} applied to the potential perturbations of Section \ref{od}.\\

Although the second order at least for potential pertubations nicely reduces to time derivatives of equilibrium correlations, it is less ready for interpretation than the first order. The expressions obtained from the path integral formalism work on a more generic level, which helps in identifying the unifying features of second order response, and serve as a more appropriate companion to the Kubo formula.

\subsection{Biased diffusion}\label{irrt}

We make here a perturbative calculation in the case of a biased diffusion which is much simpler than but effectively imitates the situation of example \ref{diel}.

Our illustration uses the following diffusion relaxational dynamics for a density profile $\rho(x,t)$ on the interval $[0,1]$:
\[
\frac {\partial \rho}{\partial t} =
 (q - p) \frac{\del \rho}{\del x} + \frac 12 (p + q) \frac{\del^2 \rho}{\del x^2} \]
We also impose the boundary condition that there is no particle current at the boundaries: $(p - q)  \rho(0,t) - \frac 12 (p + q) \rho'(0,t) = 0 = (p - q)  \rho(1,t) - \frac 12 (p + q) \rho'(1,t)$.  All that simply represents the evolution of the density
of independent particles in a closed volume and subject to an external field $E = \log \frac{p}{q}$. To that external driving there is a compensation from the diffusive current to reach asymptotically in time the equilibrium exponential profile $\rho_{\text{eq}}(x) \propto \exp[2\frac{p-q}{p+q}\,x ]$ having zero current $(p-q)\rho_{\text{eq}}(x) - \frac 1{2} (p+q) \rho'_{\text{eq}}(x) =0$ everywhere.  When the field gets changed, $E\rightarrow E_t := E+ \varepsilon\,a_t$, we should adjust the $p$ and $q$ in our mathematical description, first via
\[
h := \frac{p}{q} \longrightarrow  h_t = \frac {p_t}{q_t} = e^{E_t} = h\,(1 + \varepsilon a_t) 
\]
which is maintaining detailed balance for any fixed time with $a_t$ parameterizing the excess entropy flux, but also possibly via the symmetric quantity
\[
g := p+ q\longrightarrow g_t  = p_t + q_t=g\,(1+ \varepsilon \phi_t)
\]
which represents the modification of the escape rate of the particles.  {\it A priori}  $\phi_t$ is some unknown function of $E$ and $a_t$ (and depending in other cases also on geometry, interactions etc.).   We want to check at what order the  $\phi_t$ enters the perturbed solution.\\
We now have 
a driven diffusion with time-dependent coefficients,
\begin{equation}\label{eq:FP}
\frac {\partial \rho}{\partial t} = (q_t - p_t) \frac{\del \rho}{\del x} + \frac 12 (p_t + q_t) \frac{\del^2 \rho}{\del x^2}  \end{equation}
with
\begin{eqnarray}
p_t - q_t &=& p - q\, + \varepsilon \left[ (p -q) \phi_t + \frac{2 g\, h}{(1+h)^2}\, a_t \right]\nonumber\\
p_t + q_t &=& p + q\, + \varepsilon\,g \phi_t
\end{eqnarray}
The no-current boundary condition remains effective
so that no particles can escape from the interval. Substituting in \eqref{eq:FP} a  perturbative solution of the form, to second order,
\bea \label{exab}
\rho(x,t) = \rho_{\text{eq}}(x) + \varepsilon\, \rho_1(x,t) + \varepsilon^2 \,\rho_2(x,t)
\eea
 and comparing coefficients of powers of $\varepsilon$, we get 
to  first order  that
\[
 \frac{\del \rho_1}{\del t} = (q -p) \frac{\del \rho_1}{\del x} + \frac {g} 2 \frac{\del^2 \rho_1}{\del x^2} +  a_t \frac{2 g\, h}{(1+h)^2} \frac{\del \rho_{\text{eq}}}{\del x}
 \]
and the boundary condition reads $(q-p) \rho_1 + (p+q)\rho_1'/2 + a_t f \rho_{\text{eq}} = 0$ at $x=0,1$. Observe that this first order correction is independent of $\phi_t$; the change of the escape rates (or of the diffusion constant) is invisible at first order.  On the other hand, the second order equation at the bulk reads,
\bea
 \frac{\del \rho_2}{\del t} &=& (q - p) \frac{\del \rho_2}{\del x} + \frac 12 (p + q) \frac{\del^2 \rho_2}{\del x^2} + \phi_t \left[(q - p) \frac{\del \rho_1}{\del x} + \frac 12 (p + q) \frac{\del^2 \rho_1}{\del x^2} \right] + a_t f  \frac{\del \rho_1}{\del x} \quad
  \eea
with the boundary condition $(p-q)\rho_2 + \frac 12 \rho_2'+ a_t f \rho_1 - \phi_t a_t f \rho_{\text{eq}} = 0$ at $~ x=0,1.$
As is clearly visible, the second order does explicitly depend on $\phi_t$ via both the bulk equation and the  boundary condition.

\begin{figure}[t]
 \centering
 \includegraphics[width=8 cm]{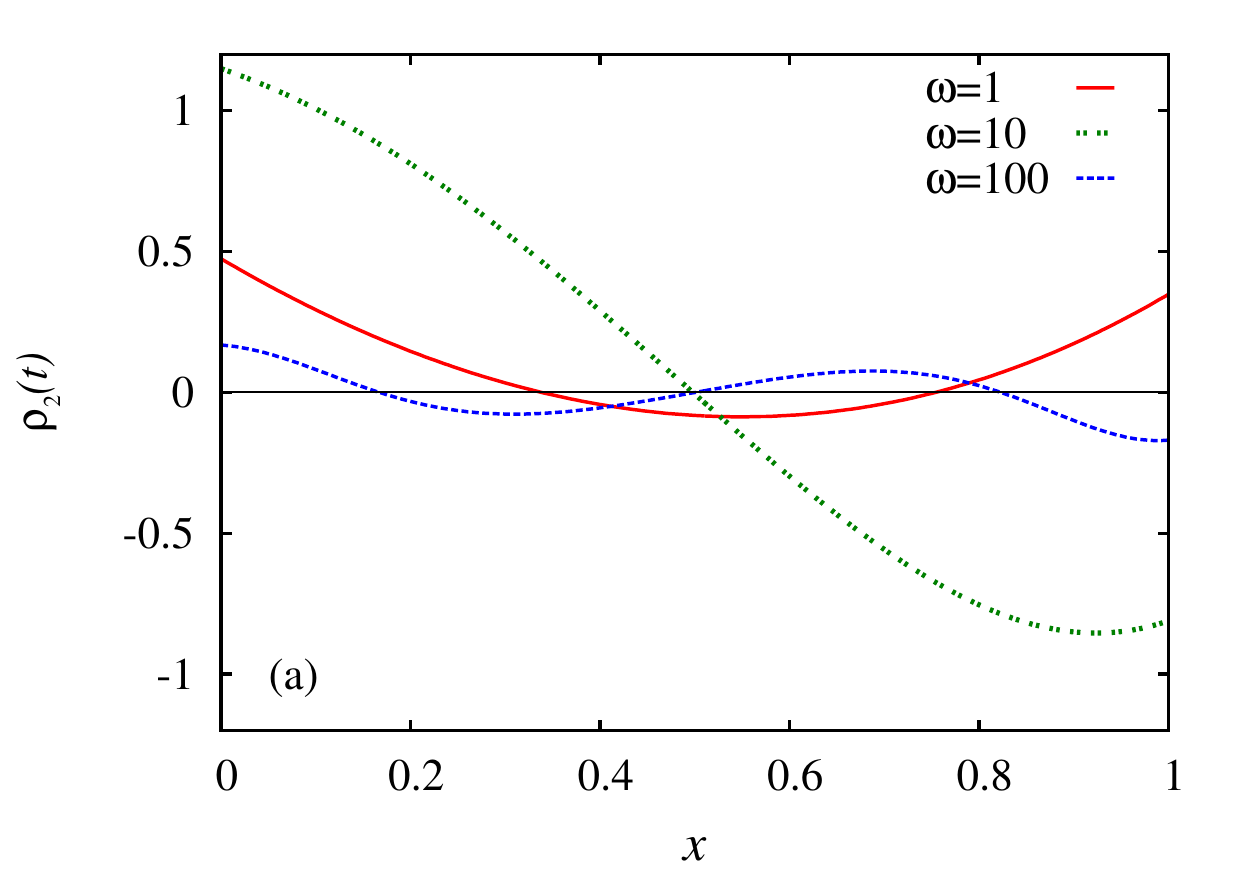} \includegraphics[width=8 cm]{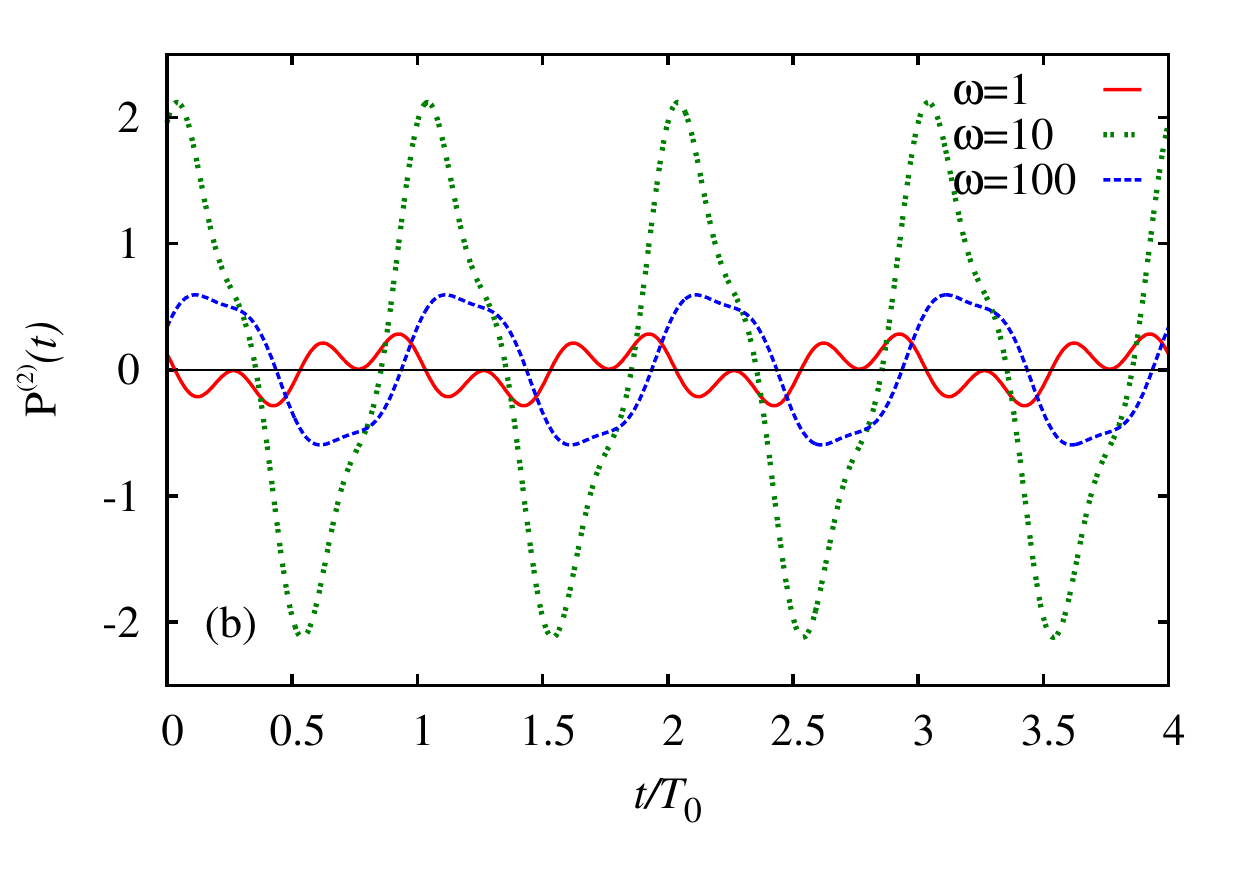}
 \caption{The asymptotic second order correction to the profile for $t=T_0/2$ (a) and second order response in the difference of boundary densities corresponding to \eqref{eq:FP} (b) for frequencies  $w =1$ (solid red), $w =10$ (dotted green) and $w= 100$ (dashed blue). Very small and very large frequencies reduce the response to zero.}
 \label{figdif}
\end{figure}

Numerical computation is summarized in Fig.\ref{figdif} for the case where $p=q=1, a_t = 2\cos w t$ and $\phi_t = -\alpha \cos^2 w t <0$ so that the escape rate diminishes with turning on the field $\varepsilon$,
\[
p_t =(1-\varepsilon\,\alpha\,\cos^2w t)(1+\varepsilon
\, \cos w t),\quad q_t =  (1-\varepsilon\,\alpha\,\cos^2 w t)(1-\varepsilon
\, \cos wt)
\]
Fig.~\ref{figdif}(a) shows the second order correction to the profile $\rho_2(x,t)$ for a fixed time $t=T_0/2$ where time period $T_0=2\pi/w.$ The second order response is zero for $\alpha=0$, and Fig.~\ref{figdif} (b) shows the periodic component in the second order response in polarization $\rho_2(1,t)-\rho_2(0,t)$ as a function of rescaled time $t/T_0$ for $\alpha=1/4$.\\

The same phenomenon also appears for nonlinear diffusions, say of the form
\[
\frac {\partial \rho}{\partial t}(\vec r,t) + \nabla \cdot j(\vec r, t) = 0, \quad j(\vec r, t) = -\chi_\varepsilon(\rho(\vec r,t))[\nabla G(\rho(\vec r,t)) - f_\varepsilon(\vec r ,t)]
\]
for a density field $\rho(\vec r,t)$ in a smooth volume with boundary $\partial V$ through which no current is allowed: $j(\vec r, t) =0$ for  $\vec r \in \partial V$.  The potential $G$ has the meaning of a local chemical potential depending on the profile $\rho$ in $V$.  In equilibrium $\nabla G(\rho_{\text{eq}}(\vec r,t))=0$ where $\rho_{\text{eq}}$ is the $\varepsilon=0$ unperturbed stationary solution, corresponding to mobility matrix $\chi_0$ and say force $f_0 =0 $.   If we perturb the mobility as $\chi_\varepsilon(\rho(\vec r,t))
= \chi_0 + \varepsilon\,\phi_t(\rho(\vec r,t))$ keeping its positivity, and the force $f_\varepsilon = \varepsilon\, a_t$, then a simple first order calculation in $\varepsilon$ shows that the first order $\rho_1$ does not depend on $\phi_t$, but the second order $\rho_2$ will, similarly to the example above with \eqref{exab}.  


%
%

\end{document}